\documentclass[preprint,prb]{revtex4}
\usepackage{graphicx}
\usepackage{latexsym}
\usepackage{amsfonts}
\begin{document}

\title{The geometrical basis of the non-linear gauge}

\author{Jose A. Magpantay}
\email{jamag@nip.upd.edu.ph}

\affiliation{ National Institute of Physics, University of the
Philippines, Diliman Quezon City, 1101,Philippines }

\date{\today}

\begin{abstract}
We consider Yang-Mills theory in Euclidean space-time $(R^4)$ and
construct its configuration space.  The orbits are first shown to
form a congruence set.  Then we discuss the orthogonal gauge
condition in Abelian theory and show that Coulomb-like surfaces
foliate the entire configuration space.  In the non-Abelian case,
where these exists no global orthogonal gauge, we derive the
non-linear gauge proposed previously by the author by modifying
the orthogonality condition.  However, unlike the Abelian case,
the entire configuration space cannot be foliated by submanifolds
defined by the non-linear gauge.  The foliation is only limited to
the non-perturbative regime of Yang-Mills theory.
\end{abstract}

\maketitle

\section{Introduction}

Today, a mathematical treatment of Yang-Mills theory generally
makes use of fiber bundles and topology\cite{top}.  But in spite
of the use of such powerful mathematics, we are nowhere near the
solution to the problem of confinement.  In fact, however, this
statement is not exactly correct.  Physicists do catch a glimpse
of confinement by making use of particular gauge conditions.  Some
examples are the Abelian\cite{gauge}, center\cite{cond} and
non-linear gauges\cite{linear}.  These gauge conditions focus on
specific configurations -monople for Abelian, vortex for center
and spherically synemetric scalars $(f^a=\partial \cdot{A}^a)$ for
the non-linear gauge, which may be responsible for confinement.

Naively, this result is paradoxical because confinement seems to
be dependent on the choice of gauge.  Gauge theorists have always
assumed that physical phenomena are gauge-independent.  But is
this really true?  In electrodynamics and perturbative non-Abelian
theory, the equivalence of quantization in various linear gauges
can be shown using formal operations on the path-integral.
Alternatively, in a particular gauge, gauge-invariance is
guaranteed by the Ward-Takahashi identity for Abelian theory and
Lee-Slavnov identities for non-Abelian theory.

However, it is also true that physical states of the gauge fields
are more transparent in certain gauges.  For example, in Abelian
theory and in the short-distance regime of the non-Abelian theory,
the transverse photon and gluons satisfy the Coulomb gauge.  This
shows that an appropriate choice of gauge can expose the physical
degrees of freedom.  Thus, if confinement is due to a specific
gauge field or a class of gauge fields, then choosing a gauge,
which highlights the field configuration(s) is absolutely
necessary.

The gauge-independence of physical results must only be true then
for gauge-fixing conditions that intersect all the orbits.  This
will guarantee that all field configurations are represented in
the path-integral.  Thus, if certain physical phenomena are
transparent in one gauge, the same physical phenomena must also be
accounted for, although may not be as transparent, in another
gauge as long as the two gauge conditions intersect all the
orbits.

In this paper, we will discuss the problem of gauge-fixing by
analyzing the configuration space of Yang-Mills theory.  We will
be employing concepts used in finite dimensional Euclidean space
and extend them in the infinite dimensional configuration space.
To visualize the concepts used, we will $\textbf{naively}$ count
the dimension and the number of elements in the gauge parameter
and configuration spaces of both the Abelian and non-Abelian
theories.  We then show that a global orthogonal gauge can be
defined for the Abelian theory but not for the non-Abelian case.
Next we present arguments why non-linear gauge-fixing is natural
for the Yang-Mills theory.  Then we modify the orthogonality
condition to derive the non-linear gauge.  Unfortunately, unlike
in the Abelian theory where Coulomb-like sufaces foliate the
entire configuration space, the non-linear submanifolds seem to be
valid only in the non-perturbative regime.

\section{The Geometry of Configuration Space}

Consider Yang-Mills theory in 4D Euclidean space-time.  The configuration space is an infinite dimensional space where the (Cartesian) axes are $A^a_\mu(x)$, i.e., the components of the gauge field at each point defined by $x_\alpha, \alpha=1,2,3,4$.  The dimension of the configuration space is $\mathcal{N} = 3\times 4\times (2N)^{4}$, where 3 comes from the SU(2) index $a$, 4 from the Lorentz index $\mu$ and $(2N)^{4}$ from the Euclidean space-time coordinates (the number of points on a line is $2N$, where $N$ is very large and approaches $\infty$ as the spacing $\epsilon$ between points approaches zero). The configuration space can then be viewed as $\mathcal{N}=4\times 3\times (2N)^{4}$ dimensional.

In configuration space a gauge field function $A^a_\mu(x)=a^a_\mu(x)$ is just a point.  We can also treat this as a ``vector'', which is pictorially represented by connecting the origin $\mathcal{A}=0$ (with components $A^{a}_{\mu}(x)=0$) to the point $\mathcal{A}=a$ (with components $A^{a}_{\mu}(x)=a^{a}_{\mu}(x)$). This ``vector'' can also be represented by a $(\mathcal{N}\times1)$ column vector $\mathit{a}$ the components of which are the values of $a^a_\mu(x)$ (all real) for each $a$, $\mu,$ and x.  The configuration space is flat as reflected by the norm
\begin{equation}\label{1}
\Vert\mathit{a}\Vert^2=\int d^4x  a^a_\mu(x) a^a_\mu(x).
\end{equation}
This means that the ``metric'' in configuration space is $\delta^{ab}\delta_{\mu\nu}\delta^4(x-x')$.\\

In the following we will only consider square-integrable $(L^{2})$ fields.  This means that the configuration space is not $R^{\mathcal{N}}$ but $B^{\mathcal{N}}$ with maximum ``radius'' $\mathcal{L}=\Vert\mathcal{A}\Vert_{max}$.  The ``volume'' of the configuration space, which counts the number of $L^{2}$ fields is $V\sim \mathcal{L}^{\mathcal{N}}$\\
The gauge transformation, which leaves the Yang-Mills action
\begin{equation}\label{2}
S=\frac{1}{4}\int{d^4}xF^a_{\mu\nu}(x)F^a_{\mu\nu}(x),
\end{equation}
invariant is
\begin{eqnarray}\label{3}
A^{'}_{\mu}&=&\Omega A_{\mu}\Omega^{-1}-i(\partial_{\mu}\Omega)\Omega^{-1}\nonumber\\
&=&\Omega[A_{\mu}+i(\partial_{\mu}\Omega^{-1})\Omega]\Omega^{-1},
\end{eqnarray}
where
\begin{eqnarray}
\Omega&=&exp[{i\wedge}],\label{4}\\
\wedge &=&\vec{\wedge}\cdot{T}=\wedge^{a}(x)T^{a},\label{5}
\end{eqnarray}
is an element of SU(2).  Using
\begin{eqnarray}
\Omega A_{\mu}\Omega^{-1}&=& A_{\mu}+i[\wedge,A_{\mu}]-\frac{1}{2}[\wedge,[\wedge,A_{\mu}]]+\cdots,\label{6}\\
-i(\partial_{\mu}\Omega)\Omega^{-1}&=&\partial_{\mu}\wedge+\frac{1}{2}[\wedge,\partial_{\mu}\wedge]-\frac{1}{6}[\wedge,[\wedge,\partial_{\mu}\wedge]]+\cdots,
\label{7}
\end{eqnarray}
the gauge transformation can be written in configuration space as
\begin{eqnarray}
\mathcal{A}^{\prime}&=&R_{\Omega}\mathcal{A}+T_{\Omega},\label{8}\\
&=& R_{\Omega}(\mathcal{A}-T_{\Omega^{-1}}).\label{9}
\end{eqnarray}
$R_{\Omega}$ is $(\mathcal{N}\times\mathcal{N})$ and its action on $\mathcal{A}$ is given by
\begin{eqnarray}
[R_{\Omega}\mathcal{A}]^a_{\mu}(x)=2\int{d^{4}}x'\delta_{\nu\mu}\delta^{4}(x-
x')tr\{T^{a}\langle1+i[\wedge, \quad]-\frac{1}{2}[\wedge,[\wedge,    ]]\nonumber\\
-\frac{1}{6}[\wedge,[\wedge,[\wedge,   ]]]+\cdots\rangle{A}_{\nu}(x')\}.\label{10}
\end{eqnarray}
$T_{\Omega}$, on the other hand is $(\mathcal{N}\times 1)$ and its components are read from Equation (\ref{7}).

Since $\parallel\Omega \mathcal{A}\Omega^{-1}\parallel=\parallel \mathcal{A}\parallel$, then
\begin{eqnarray}
R^+_{\Omega} R_{\Omega}&=&\mathbf{1}    \quad(\mathcal{N}\times{\mathcal{N}}\quad identity)\nonumber\\
&=& \mathbf{1}_{SU(2)}\otimes\mathbf{1}_{Lorentz}\otimes\delta^{4}(x-x')\label{11}.
\end{eqnarray}
We will take $det\ R_{\Omega}=1$. Equations (\ref{8}), (\ref{9}) and (\ref{11}) establish that gauge transformation is a combination of translation and rotation in configuration space.  This makes the configuration space an affine space.

The gauge parameter $(\Omega)$ space is $\mathcal{D}=3\times(2N)^{4}$ dimensional.  Since we will require $\Vert\mathcal{A}^{\Omega}\Vert < \mathcal{L}$, then the gauge parameter space must be a Sobolev space\cite{chodos}.  In this space, $(\Omega-{1})$ is continuously differentiable and an element of the Hilbert space with norm
\begin{equation}\label{12}
\Vert\Omega-1\Vert^{2} = \int\frac{d^{4}x}{x^{2}}tr[(\Omega-1)^{\dagger}(\Omega-1)]+\int d^{4}x\ tr[\partial_{\mu}\Omega^{-\dagger}\partial_{\mu}\Omega].
\end{equation}
For this norm to be finite, we exclude constant gauge transformation except $\Omega=\textbf{1}$.

Let us now focus on pure gauge fields
\begin{equation}\label{13}
a_{\mu}=-i(\partial_{\mu}\Omega)\Omega^{-1}.
\end{equation}
with $\Omega\  \epsilon\ {SU(2)}$.  Note that this field configuration has zero field strength; thus it contains the trivial vacuum.  Equation (13) maps the $\mathcal{D}$ dimensional gauge parameter space (Sobolev completed) to corresponding points in the configuration space $B^{\mathcal{N}}$.  Furthermore, each pure gauge $a_{\mu}$ corresponds to a unique $\Omega$.  This easily follows from if
\begin{equation}\label{14}
\Omega^{-1}\partial_{\mu}\Omega = \tilde{\Omega}^{-1}\partial_{\mu}\tilde{\Omega},
\end{equation}
then
\begin{equation}\label{15}
(\partial_{\mu}\Omega)=\Omega\tilde{\Omega}^{-1}\partial_{\mu}\tilde{\Omega}=(\tilde{\Omega}\Omega^{-1})^{-1}\partial_{\mu}[\tilde\Omega\Omega^{-1}\Omega].
\end{equation}
This gives $\tilde{\Omega}\Omega^{-1}\ \epsilon\ SU(2)$ must be a constant, which must be equal to identity by Sobolev completion. Equivalently, as suggested by equations (\ref{8}) and (\ref{9}), each point in the function space of the gauge parameters is mapped to $T_\Omega$, which belongs to a class of translation group which have vanishing field strengths.  Let us call this particular class of the translation group $\tau_o$.

We will now show that $\tau_o$ forms an orbit that passes through the origin of the configuration space.  Note that the origin $\mathcal{A}=0$ is uniquely determined by $\Omega=\textbf{1}$, again because of Sobolev completion. We need to show that a gauge transform of a pure gauge field is also a pure gauge field.  Let $a_{\mu}=i(\partial_{\mu}\Omega)\Omega^{-1}$ and consider its gauge transform under $\tilde{\Omega}$, i.e.,
\begin{eqnarray}
a'_{\mu}&=&\tilde{\Omega}i(\partial_{\mu}\Omega)\Omega^{-1}\tilde{\Omega}^{-1}-i(\partial_{\mu}\tilde{\Omega})\tilde{\Omega}^{-1}\nonumber\\
&=& -i[\partial_{\mu}(\tilde{\Omega}\Omega](\tilde{\Omega}\Omega)^{-1}.\label{16}
\end{eqnarray}
This shows that $a'_{\mu}$ is a pure gauge with gauge element $\Omega'=\tilde{\Omega}\Omega$.  In configuration space, equation (\ref{14}) becomes
\begin{equation}\label{17}
T'_{\Omega}=R_{\tilde{\Omega}}T_{\Omega}+T_{\tilde{\Omega}}=T_{\Omega'}
\end{equation}
Equations (\ref{16}) and (\ref{17}) show that we can generate $\tau_o$, the orbit of the pure gauge configurations which has zero field strength, from the origin $\mathcal{A}=0.$

Let the orbit passing through a point $\mathcal{A}$ in configuration space be $\tau_{\mathcal{A}}$.  This can be generated in two ways.  The first is by rotating the vector $\mathcal{A}$ using $R_{\Omega}$ and then translating with $T_{\Omega}$ as presented in equation (\ref{8}).  This should be done on $\mathcal{A}$ using all the operations $\{R_{\Omega},T_{\Omega}\}$.  The second is by first translating $\mathcal{A}$ by $(-)T_{\Omega^{-1}}$ and then rotating by $R_{\Omega}$.  This is prescribed in equation (\ref{9}) and should also be done using all the operations $\{R_{\Omega},T_{\Omega^{-1}}\}$.

\begin{figure}
\centering
\includegraphics[totalheight=2.5in]{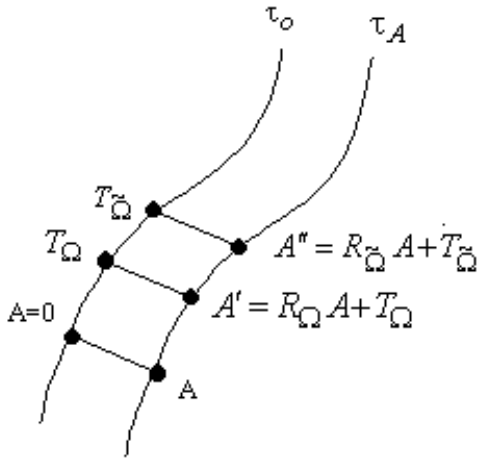}
\caption{The orbits $\tau_{o}$ and $\tau_{A}$ showing three sets of corresponding points connected by steps of a ladder.  The distances between each set of points, which is equal to the length of the ladder steps, are all equal to $\Vert\mathcal{A}-0\Vert$.}
\label{fig1}
\end{figure}

\begin{figure}
\centering
\includegraphics[totalheight=2.5 in]{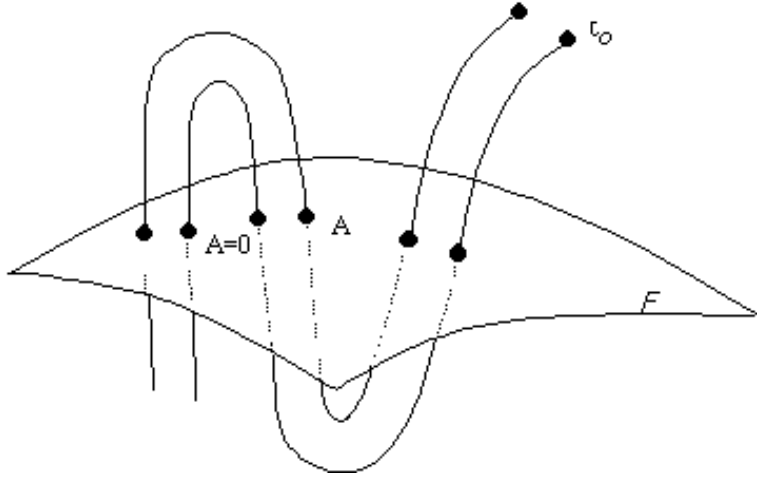}
\caption{The gauge-fixing surface $\mathcal{F}$, which is intersected by the orbits $\tau_{o}$ and $\tau_{A}$ at three points each.  Note the distance between corresponding points on the orbits are all the same}.
\label{fig2}
\end{figure}

The orbits $\tau_{\mathcal{A}}$ and $\tau_o$, no matter how they twist and turn in configuration space, always maintain the distance $\parallel{\mathcal{A}}\parallel$ between their corresponding points as shown in Figure \ref{fig1}.
This ladder-like structure follows from applying equation (\ref{8}) on both $\mathcal{A}$ and the origin and then subtracting the gauge transformed results.  This will yield $R_{\Omega}(\mathcal{A}-0)$, which has the same norm as $\mathcal{A}$.  Doing this for all $\{R_{\Omega},T_{\Omega}\}$ we generate the ladder-like structure of the two orbits.  If this is true for $\tau_{\mathcal{A}}$ and $\tau_o$, it is also true for $\tau_{\tilde\mathcal{A}}$ and $\tau_o$ and also for $\tau_{\mathcal{A}}$ and $\tau_{\tilde{\mathcal{A}}}$.  The orbits, though twisting and turning in a complicated way, maintain the same distance between corresponding  points on the two orbits.  The orbits therefore form a congruence set, i.e., they cover the configuration space without intersecting.  This simple observation is significant for the following reasons.  If the gauge-fixing submanifold does not intersect $\tau_o$ uniquely, it will not intersect neighboring orbits uniquely also (see Figure \ref{fig2}).  If we know how $\tau_o$ twists and turns, then we know how all the other orbits twist and turn also.  This will suggest how to choose the gauge-fixing submanifold, if not throughout the entire configuration space, at least in the vicinity of physically interesting field configurations.

Finally, we note why the path-integral is invariant under gauge transformation.  The path-integral measure
\begin{equation}\label{18}
[dA^{a}_{\mu}(x)]={\prod_{a,\mu,x_{\alpha}}dA^{a}_{\mu}(x_{\alpha})={\prod_{I=1}^{\mathcal{N}}dA_{I}}}
\end{equation}
where the last term is the ``infinitesimal volume'' in configuration space.  This measure is invariant under the affine transformation defined by equation (\ref{8}).

The gauge-invariant action can be written as
\begin{eqnarray}
S=-\frac{1}{2}\int d^{4}x d^{4}x'A^{a}_{\mu}(x')\{\delta^{4}(x-x')\delta^{ab}[\Box^{2}\delta_{\nu\mu}-\partial_{\mu}\partial_{\nu}]\}A^{b}_{\nu}(x)\nonumber\\
+\int d^{4} x d^{4}x'd^{4}x''A^{b}_{\mu}(x')A^{c}_{\nu}(x'')\{\delta^{4}(x-x')\delta^{4}(x'-x'')\epsilon^{bca}\delta_{v\alpha}\partial_{\mu}\}A^{a}_{\alpha}(x)\nonumber\\
+\frac{1}{2}\int d^{4}xd^{4}x'd^{4}x''d^{4}x'''\{[\delta^{ac}\delta^{bd}-\delta^{ad}\delta^{bc}]\delta^{4}(x-x')\delta^{4}(x'-x'')\nonumber\\
\times\delta^{4}(x''-x''')\delta_{\mu\alpha}\delta_{\nu\beta}\}A^{a}_{\mu}(x)A^{b}_{\nu}(x')A^{c}_{\alpha}(x'')A^{d}_{\beta}(x''')\label{19}
\end{eqnarray}

The action forms a quartic hyperplane (we use the convention where hyperplane is a submanifold of dimension one less than the manifold) in configuration space of the general form
\begin{equation}\label{20}
S=\alpha_{IJ}\mathcal{A}_{I}\mathcal{A}_{J}+\beta_{IJK}\mathcal{A}_{I}\mathcal{A}_{J}\mathcal{A}_{K}+\gamma_{IJKL}\mathcal{A}_{I}\mathcal{A}_{J}\mathcal{A}_{K}\mathcal{A}_{L}.
\end{equation}
Although it is not apparent, this form of the action is invariant under the combined operations of rotation and translation as given by equation (\ref{9}).  Equations (\ref{18}) and (\ref{19}) establish the invariance of the path-integral under gauge-transformation.

The normal vector to the hyperplane S=constant at $\mathcal{A}$ is $\mathcal{P}$ with components
\begin{equation}\label{21}
P^{a}_{\mu}(x)=\frac{\delta{S}}{\delta{A^{a}_{\mu}(x)}}=D^{ab}_{\nu}F^{b}_{\nu\mu}(x).
\end{equation}
The tangent to the orbit at $\mathcal{A}$ is $\mathcal{T}$ with components
\begin{equation}\label{22}
T^{a}_{\mu}(x)=D^{ab}_{\mu}\Lambda^{b}.
\end{equation}
The expected orthogonality of $\mathcal{P}$ and $\mathcal{T}$ follow from
\begin{eqnarray}
\mathcal{P}\cdot\mathcal{T}&=&\int{d^{4}}x(D^{ab}_{\nu}F^{b}_{\nu\mu})(D^{ac}_{\mu}\Lambda^{c})\nonumber\\
&=&-\int{d^{4}}x(D^{ca}_{\mu}D^{ab}_{\nu}F^{b}_{\nu\mu})\Lambda^{c}\nonumber\\
&=&0.\label{23}
\end{eqnarray}

\section{The Orthogonal Gauge Condition}

Gauge fixing is the process of choosing representative field configurations from each orbit.  Ideally, the gauge-fixing should choose only one representative from each orbit and all orbits should be represented.  This condition is equivalent to saying that the gauge-fixing condition is unique and always realizable.

Generally, gauge-fixing is done by imposing a local condition on the potentials, i.e., requiring
\begin{equation}\label{24}
F^{a}[A_{\mu}(x)]=0,\quad{a}=1,2,3.
\end{equation}
Uniqueness requires that if $A_{\mu}(x)$ satisfies equation (\ref{22}), then $F^{a}[A^{\Omega}_{\mu}(x)]\neq{0}$, i.e., all the gauge transformed fields of $A_{\mu}(x)$ must not satisfy the condition.  Realizability requires that for all $A_{\mu}(x)$  which does not satisfy equation (\ref{24}), there must be an $\Omega\  \epsilon\ {SU(2)}$ such that $F^{a}[A^{\Omega}]=0$.

The ideal condition is satisfied in only one case, the Coulomb gauge fixing of an Abelian theory.  In all the other linear gauge-fixing of Abelian and non-Abelian theories, realizability is generally taken for granted while non-uniqueness is rectified through subsidiary conditions (Gupta-Bleuler condition in Lorentz gauge) or Fadeev-Popov determinants.

In configuration space, gauge-fixing is tantamount to choosing a submanifold where all orbits must pass through.  Since one of the four $A^{a}_{\mu}(x)$ for each a and at each x essentially becomes a dependent variable when $F^{a}[A]=0$ is imposed, the submanifold $\mathcal{F}$ defined by the gauge-fixing is naively $3\times3\times(2N)^{4}$ dimensional.  The issue now is what geometrical principle should be used in choosing $\mathcal{F}$.  The simplest and most compelling principle is to require global orthogonality of the orbit to $\mathcal{F}$.  This will guarantee uniqueness and realizability.  As stated already this is only achieved in the Coulomb gauge formulation of an Abelian theory.  This will be shown below.

In 4D euclidean space, the Coulomb gauge is given by the local condition
\begin{equation}\label{25}
F[A]=\partial_{\mu}A_{\mu}=0.
\end{equation}
For an Abelian theory, uniqueness and realizability of this condition follow from the positive definiteness of the Laplacian operator.  The configuration space, naively, has dimension $\mathcal{N}=4\times(2N)^{4}$.  From equation (\ref{25}), there are only three independent potentials at each point thus the submanifold of transverse potentials $\mathcal{F}$ in configuration space has dimension $\mathcal{M}=3\times(2N)^{4}$.  This means that $\mathcal{F}$ is defined by
\begin{equation}\label{26}
\mathcal{F}_{c_{I}\cdots c_{\mathcal{H}}}=F^{\Lambda_1}\cap{F^{\Lambda_2}}\cap\cdots,F^{\Lambda_\mathcal{H}},
\end{equation}
i.e., it is the intersection of hyperplanes (dimension equal to $4\times(2N)^{4}-1$) defined by each $F^{\Lambda_{I}}=c_{I}$, the total of which is $\mathcal{H}=(2N)^{4}$.  Below, we will determine $F^{\Lambda}$ geometrically.  We will also argue that there is a sufficient number of $\Lambda$'s in the U(1) gauge parameter space to ensure that $\mathcal{F}$ is an appropriate submanifold.\\
Let us now determine $F^{\Lambda}$ by imposing that the normal to $F^{\Lambda}$ at $\mathcal{A}$ is equal to the tangent to the orbit.  This is equivalent to the orbit being orthogonal to the $F^{\Lambda}$=const. hyperplane.  In component form, this condition means
\begin{equation}\label{27}
\frac{\delta F^{\Lambda}}{\delta A_{\mu}(x)}=\partial_{\mu}\Lambda(x).
\end{equation}

The solution is
\begin{eqnarray}
F^{\Lambda}&=&\int{d^{4}}x(\partial_{\mu}\Lambda(x))A_{\mu}(x)\nonumber\\
&=&(-)\int{d^{4}}x\Lambda(x)(\partial_{\mu}A_{\mu}(x))\label{28}
\end{eqnarray}
Imposing all $c_{I}$'s equal to zero gives the local condition $\partial_{\mu}A_{\mu}=0$, i.e., the Coulomb gauge, while $c_I\neq 0$ corresponds to $\partial_{\mu}A_{\mu}=f(x)$.\\

At this point, all we have shown is that we can find a hyperplane $F^{\Lambda}$ such that its normal is parallel to the tangent to the orbit.  But the gauge-fixed submanifold $\mathcal{F}$, which is $3\times(2N)^{4}$ dimensional is defined as the intersection of $(2N)^{4}$ hyperplanes.  Can we find $(2N)^{4}$ gauge parameters, which define the hyperplanes such that their intersections define a $3\times(2N)^{4}$ dimensional $\mathcal{F}$.  This means no two $F^{\Lambda^{\prime} s}$ should have parallel normal vectors.  Imposing that the normal to $F^{\Lambda}$ and $F^{\prime}$ are orthogonal, we find
\begin{eqnarray}\label{29}
\frac{\delta F^{\Lambda}}{\delta \mathcal{A}}\cdot\frac{\delta F^{\Lambda^{\prime}}}{\delta \mathcal{A}}&=&\int d^{4}x\ \partial_{\mu}\Lambda(x)\ \partial_{\mu}\Lambda^{\prime}(x)\nonumber\\
&=& -\int d^{4}x\ \Lambda(x)\Box^{2}\Lambda^{\prime}(x).
\end{eqnarray}
The vanishing of the surface term follow from the fact that since we only consider $L^{2}$ gauge fields, then under gauge transformation $A^{\prime}_{\mu}=A_{\mu}+\partial_{\mu}\Lambda$, the triangle inequality says.
\begin{equation}\label{30}
[\int d^{4}x\ (\partial_{\mu}\Lambda)^{2}]^{\frac{1}{2}}< \Vert \mathcal{A}\Vert + \Vert \mathcal{A}^{\prime}\Vert < \infty.
\end{equation}
Equation (30) implies that $\Lambda$ must have suitable behaviour at $\infty$ (goes to zero faster than $\frac{1}{r}$ at $\infty$).

Since $\Box^{2}$ is a Hermitean, positive definite operator, equation (29) implies
\begin{equation}\label{31}
(\lambda -\lambda^{\prime})\int d^{4}x\ \Lambda(x)\ \Lambda^{\prime}(x) = 0,
\end{equation}
where $\lambda$ and $\lambda^{\prime}$ are the eigenvalues of $\Lambda$ and $\Lambda^{\prime}$ under the action of $\Box^{2}$.  We now argue that there is a sufficient number of $\Lambda$'s (at least $(2N)^{4})$ that are eigenfunctions of $\Box^{2}$ with different eigenvalues.  These have the form
\begin{equation}\label{32}
\Lambda(x)=\int d^{4}k\ K(k)e^{ik\cdot x}.
\end{equation}
Imposing $\Lambda$ satisfies $\Box^{2}\Lambda=\lambda\Lambda(x)$, where $\lambda > 0$, $\Lambda$ becomes
\begin{equation}\label{33}
\Lambda(x)=\int d^{3}k\ K(\vec{k}, k_{4}=(\lambda-\vec{k}\cdot \vec{k})^{\frac{1}{2}})e^{i\vec{k}\cdot\vec{x}+i(\lambda-\vec{k}\cdot\vec{k})^{\frac{1}{2}}x_{4}}.
\end{equation}
To satisfy equation (\ref{30}), $K(\vec{k}, k_{4}=(\lambda-\vec{k}\cdot\vec{k})^{\frac{1}{2}})$ must satisfy
\begin{equation}\label{34}
\int d^{3}k\ K^{2}(\vec{k}, k_{4}=(\lambda-\vec{k}\cdot\vec{k})^{\frac{1}{2}}) < \infty.
\end{equation}
This means $K$ goes to zero faster than $\frac{1}{\vert\vec{k}\vert^{1\cdot 5+\epsilon}}$ as $\vert\vec{k}\vert\rightarrow\infty$.  Definitely there are infinitely many functions with such behaviour.  This proves that we can find a sufficient number of $\Lambda$'s that can define $F^{\Lambda}$.

Equation (\ref{31}) implies that equation (\ref{29}) gives zero.  This means that we can choose $(2N)^{4}$ hyperplanes $F^{\Lambda^{\prime}_{i}}$ with $i=1,\cdots,(2N)^{4}$, such that all their normals are orthogonal to each other.  This shows that the submanifolds $\mathcal{F}_{c_{1}\cdots c_{\mathcal{H}}}$ foliate the configuration space.

Equivalently, we can show that the submanifolds defined by equations (\ref{28}) and (\ref{26}) foliate the configuration space by making use of the Frobenius theorem.  In the following, we will use the form version.  Define the set of one-forms $((2N)^{4}$ in total) in configuration space, which live on the cotangent space, by\\
\begin{equation}\label{35}
w^{\Lambda}=\int{d^{4}}x(\partial_{\mu}\Lambda)dA_{\mu}(x).
\end{equation}
Since $w^{\Lambda}=dF^{\Lambda}$, the set of one forms is a closed set.  The ``new coordinates'' $F^{\Lambda}$ defined by equation (\ref{28}) form a surface $\mathcal{F}_{c_{I}\cdots c_{\mathcal{H}}}$ given by equation (\ref{26}) when each $F^{\Lambda_{I}} = c_{I}$, for $I = 1,\cdots\mathcal{H}$.

From this construction, it follows that
\begin{equation}\label{36}
w^{\Lambda
}\vert_{\mathcal{F}_{_{c_{I}}\cdots c_{\mathcal{H}}}}= 0,
\end{equation}
i.e., on the submanifold $\mathcal{F}_{c_{I}\cdots c_{\mathcal{H}}}$, the tangent vectors annul the one forms.  In particular, on the Coulomb gauge submanifold given by $\mathcal{F}_{o \cdots o}$, this follows from

\begin{eqnarray}
w^{\Lambda}\vert_{\mathcal{F}_{_{o \cdots o}}}&=&\int d^{4}x \partial_{\mu}\Lambda dA_{\mu}(x)\vert_{\mathcal{F}_{_{o \cdots o}}}\nonumber\\
&=&(-)\int{d^{4}}x\Lambda(x)\partial_{\mu}dA_{\mu}(x)\vert_{\mathcal{F}_{_{o\cdots o}}}.\label{37}
\end{eqnarray}
Since $A_{\mu}(x)$ is transverse on $\mathcal{F}_{o\cdots o}$, we can write
\begin{equation}\label{38}
A_{\mu}(x)\vert_{\mathcal{F}_{_{o\cdots o}}} = (\delta_{\mu\nu}-\partial_{\mu}\frac{1}{\Box^{2}}\partial_{\nu})A_{\nu},
\end{equation}
giving
\begin{equation}\label{39}
dA_{\mu}(x)\vert_{\mathcal{F}_{_{o\cdots o}}} = (\delta_{\mu\nu}-\partial_{\mu}\frac{1}{\Box^{2}}\partial_{\nu})dA_{\nu}.
\end{equation}

Substituting equation (\ref{39}) in equation (\ref{34}) verifies equation (\ref{36}). All these prove that the submanifolds $\mathcal{F}_{c_{I}\cdots c_{\mathcal{H}}}$ where the orbits are orthogonal, are leaves in the foliation of the entire configuration space.\\

In the non-Abelian case, the tangent to the orbit is the vector $\mathcal{T}$ defined by equation(\ref{22}).  Imposing that this is equal to the normal to the surface defined by $F^{\vec{\Lambda}}$ = constant implies
\begin{equation}\label{40}
\frac{\delta F^{\vec\Lambda}}{\delta A^{a}_{\mu}(x)}=D^{ab}_{\mu}\Lambda^{b}=\partial_{\mu}\Lambda^{a}-\epsilon^{abc}A^{c}_{\mu}\Lambda^{b}.
\end{equation}
Because of the second term, there is no solution to equation (\ref{40}).  Hence, the conclusion that there exists no orthogonal gauge condition, local or global, in the non-Abelian case.  This result had been established by various authors, including Chodos and Moncrief\cite{group},
who used the vector version of Frobenius theorem.\\

\section{Geometry of the Non-Linear Gauge}

We will now discuss the non-linear gauge, which was discussed by the author in a series of articles.  Initially, the author's justification for the gauge condition is the fact that there are field configurations missed by the Coulomb gauge\cite{hadron}. These are the field configurations that are on the Gribov horizon of the $\partial\cdot{A}={\mathit{f}}\neq 0$ surface.  In subsequent papers, the author showed that the gauge condition ``reveals'' the physical degrees of  freedom, which depend on the distance scale, of the non-Abelian theory.  At short distance, i.e., well inside hadrons, transverse gluons exist and interact very weakly with quarks.  This is accounted for by the linear limit of the non-linear gauge.  At large distance scales, the important field configurations are the new scalar fields $\mathit{f}^{a}=\partial\cdot{A}^{a}$, which has an infinitely non-linear effective action.  The classical, stochastic dynamics of spherically symmetric $\mathit{f}^{a}$ leads to the linear potential\cite{linear} while the full quantum dynamics leads to dimensional reduction\cite{chaos}.

What we would like to raise at this point is the question, Is there a geometrical basis for the non-linear gauge?  The answer is yes and the arguments essentially follow equations (26) to (39) in the Abelian case.

Before we answer this question, we note that since the orbit through $\mathcal{A}$ twists and turns in configuration space (see discussions in Section II), it is most unlikely that there exists a linear submanifold that intersects all orbits uniquely.  As shown in reference (7), it may also happen that a linear submanifold may not intersect some orbits at all.  For this reason, there are those who proposed covering the submanifold by local patches centered around background gauge fields\cite{orbit}. This gauge fixing is essentially a collection of linear gauges.  However, beyond formal expressions for the path-integral and global expectation values for gauge-invariant quantities, this formalism has not really shown confinement.  Also, a collection of linear gauges actually suggests non-linearity of the entire submanifold.

The non-linear gauge is also hinted by equation (\ref{40}), which states that the orbit is orthogonal to the gauge-fixing surface.  Since the RHS of equation (\ref{40}) is linear in A, the hyperplane $F^{\vec{\Lambda}}$ = constant must be quadratic in A.  Unfortunately, the anti-symmetric $\epsilon^{abc}$ precludes the existence of a solution.

But suppose we modify equation (\ref{40}) to something like
\begin{equation}\label{41}
\frac{\delta F^{\vec\Lambda}}{\delta A^{a}_{\mu}(x)}=\int {d^{4}} x'h^{ab}_{\mu\nu}(x;x')(D^{bc}_{\nu}\Lambda^{c})_{x'}.
\end{equation}
Equation (\ref{41}) states that the normal to the hyperplane $F^{\vec{\Lambda}}$ = constant is a linear combination of the components of the tangent to the orbit at $\mathcal{A}$ (with components $(D^{ab}_{\mu}\Lambda^{b})_{x}$).  This means that the gauge-fixing submanifold $\mathcal{F}_{c_{I}\cdots c_{\mathcal{H}}}$ given by the intersections of the hyperplanes $F^{\vec{\Lambda_{I}}}=c_{I}$, i.e.
\begin{equation}\label{42}
\mathcal{F}_{c_{I}}\cdots c_{\mathcal{H}} =F^{\vec{\Lambda}_{1}}\cap F^{\vec{\Lambda}_{2}}\cap F^{\vec{\Lambda}_{3}}\cap\cdots,\cap F^{\Lambda_{\mathcal{H}}},
\end{equation}
intersects the orbit but is not orthogonal to it.  The submanifold $\mathcal{F}_{c_{I}\cdots c_{\mathcal{H}}}$ is tilted slightly relative to the orbit, with the tilting determined by $h^{ab}_{\mu\nu}(x;x')$ given in equation (\ref{41}).

Before we solve equation (\ref{41}), we give a naive counting of dimensions. Each $F^{\vec{\Lambda}}$ is a hyperplane (dimension equal to $3\times 4\times(2N)^{4}-1)$ and we will need a total $\mathcal{H}=3\times(2N)^{4}$ specified by choosing an equal number of $\Lambda^{a}(x)$ from the Sobolev completed gauge parameter space.  This will make the gauge-fixing submanifold $\mathcal{F}_{c_{1}\cdots c_{\mathcal{H}}}[3\times3\times(2N)^{4}]$ dimensional.

Consider the following $h^{ab}_{\mu\nu}$
\begin{equation}\label{43}
h^{ab}_{\mu\nu}(x;x')=\delta^{4}(x-x')\delta^{ab}\partial'_{\mu}\partial'_{\nu}+\frac{1}{4}\partial'_{\mu}(\partial\cdot A^{b})_{x'}\frac{\delta}{\delta A^{a}_{\nu}(x)}.
\end{equation}
Substituting in (\ref{41}), we find
\begin{equation}\label{44}
\frac{\delta F^{\vec{\Lambda}}}{\delta A^{a}_{\mu}(x)}=\partial_{\mu}(\partial\cdot D)^{ab}\Lambda^{b}+\epsilon^{abc}\partial_{\mu}(\partial\cdot A^{c})\Lambda^{b}.
\end{equation}
From equation (\ref{44}), we find
\begin{eqnarray}
F^{\vec\Lambda} &=& -\int d^{4}x\Lambda^{b}(x)[(D\cdot\partial)^{bc}(\partial\cdot A^{c})],\nonumber\\
&=&-\int d^{4}x\Lambda^{b}(x)[(\partial\cdot D)^{bc}(\partial\cdot A^{c})],\nonumber\\
&=&-\frac{1}{2}\int d^{4}x\Lambda^{b}(x)\{[(\partial\cdot D)^{bc}+(D\cdot\partial)^{bc}](\partial\cdot A^{c})\}.\label{45}
\end{eqnarray}
From equation (\ref{45}), we read that the submanifold $\mathcal{F}_{c_{I}\cdots c_{\mathcal{H}}}$ with all $c_{I} = 0$ defines the non-linear gauge condition
\begin{equation}\label{46}
(\partial\cdot D)^{ab}(\partial\cdot A^{b})=(D\cdot\partial)^{ab}(\partial\cdot A^{b})=\frac{1}{2}[(\partial\cdot D)^{ab}+(D\cdot\partial)^{ab}](\partial\cdot A^{b})= 0.
\end{equation}
And for arbitrary set of constants $c_{I}$, with $I=1,\cdots,\mathcal{H}$; the submanifold $\mathcal{F}_{c_{I}\cdots c_{\mathcal{H}}}$ is defined by the gauge condition
\begin{equation}\label{47}
(\partial\cdot D)^{ab}(\partial\cdot A^{b})=s^{a}(x),
\end{equation}
with
\begin{equation}\label{48}
F^{\vec{\Lambda}_{i}}=c_{I}=-\int d^{4}x\Lambda^{a}_{I}(x)s^{a}(x).
\end{equation}

Just like in the Abelian case, consider the dot product between the normal vectors to the $F^{\vec{\Lambda}}$ and $F^{\vec{\Lambda^{\prime}}}$ hyperplanes at $\mathcal{A}$.  This is given by
\begin{eqnarray}\label{49}
\frac{\delta F^{\vec{\Lambda}}}{\delta \mathcal{A}}\cdot\frac{\delta F^{\vec{\Lambda^{\prime}}}}{\delta \mathcal{A}}&=&\int d^{4}x\ \frac{\delta F^{\vec{\Lambda}}}{\delta A^{a}_{\mu}(x)}\frac{\delta F^{\vec{\Lambda^{\prime}}}}{\delta A^{a}_{\mu}(x)}\nonumber\\
&=&\int d^{4}x\ [\partial_{\mu}(\partial\cdot D)^{ab}\Lambda^{b}+\epsilon^{abc}\partial_{\mu}(\partial\cdot A^{c})\Lambda^{b}].\nonumber\\
&&[\partial_{\mu}(\partial\cdot D)^{ad}\Lambda^{\prime d}+\epsilon^{ade}\partial_{\mu}(\partial\cdot A^{e})\Lambda^{\prime d}].
\end{eqnarray}
Integration by parts and the vanishing of the surface terms because of the $L^{2}$ behaviour of $A_{\mu}$ and the gauge parameters belonging in Sobolev space leads to
\begin{eqnarray}\label{50}
\frac{\delta F^{\vec{\Lambda}}}{\delta \mathcal{A}}\cdot\frac{\delta F^{\vec{\Lambda^{\prime}}}}{\delta \mathcal{A}}&=&\int d^{4}x\ \Lambda^{a}(x)\Sigma^{ab}\Lambda^{\prime b}(x)\nonumber\\
&=&\int d^{4}x\ (\Sigma^{ba}\Lambda^{a}(x))\Lambda^{\prime b},
\end{eqnarray}
where the Hermitean sixth order operator $\Sigma^{ab}$ is given
\begin{eqnarray}\label{51}
\Sigma^{ab}&=&-(D\cdot\partial)^{ac}\Box^{2}(\partial\cdot D)^{cd}-\epsilon^{cbd}(D\cdot \partial)^{bc}\partial_{\mu}[\partial_{\mu}(\partial\cdot A^{e}]\nonumber\\
&+&\epsilon^{acd}(\partial_{\mu}(\partial\cdot A^{c}))\partial_{\mu}(\partial\cdot D)^{db}\nonumber\\
&+& \delta^{ab}[\partial_{\mu}(\partial\cdot A^{c})][\partial_{\mu}(\partial\cdot A^{c})]-[\partial_{\mu}(\partial\cdot A^{a})][\partial_{\mu}(\partial\cdot A^{b})].
\end{eqnarray}

Because $\Sigma^{ab}$ is a sixth-order operator that depends non-linearly on $A^{a}_{\mu}$, it is not possible to carry out an analysis that goes along the same lines as equations(\ref{32}) to (\ref{34}) to determine its eigenfunctions.  We will just argue that there must be, at least, $3\times(2N)^{4}$ eigenfunctions of $\Sigma^{ab}$ in the Sobolev completed gauge parameter space.  The reason is that this space is $3\times(2N)^{4}$ dimensional, thus the volume contains $(\delta)^{3\times(2N)^{4}}\vec{\Lambda}'s$, where $\delta$ is $\Vert\vec{\Lambda}\Vert_{max}$ in the gauge parameter space.

From (50), we find that the normal to the hyperplanes $F^{\vec\Lambda_{1}},\cdots, F^{\vec\Lambda_{\mathcal{H}}}$ can be made orthogonal to each other.  Thus, $\mathcal{F}_{c_{1}\cdots c_{\mathcal{H}}}$ defined by equation (\ref{42}) is $3\times 3\times(2N)^{4}$ dimensional and is an appropriate gauge-fixed submanifold.

Now let us consider the set of one forms
\begin{equation}\label{52}
w^{\vec{\Lambda}}=\int d^{4}x\{\partial_{\mu}(\partial\cdot D)^{ab}\Lambda^{b}+\epsilon^{abc}\partial_{\mu}(\partial\cdot A^{c})\Lambda^{b}\}dA^{a}_{\mu}(x).
\end{equation}
Just like in the Abelian case, since $w^{\vec{\Lambda}} = dF^{\vec{\Lambda}}$, the set of one forms is a closed set.  The ``new coordinates'' $F^{\vec{\Lambda}}$ form a surface $\mathcal{F}_{c_{I}\cdots c_{\mathcal{H}}}$ as given in equation (\ref{42}).\\

From this construction it follows that
\begin{equation}\label{53}
w^{\vec{\Lambda}}\vert_{\mathcal{F}_{_{c_{I}}\cdots c_{\mathcal{H}}}}= 0.
\end{equation}
And in the particular case of $\mathcal{F}_{o\cdots o}$, i.e., the submanifold defined by the nonlinear regime of the non-linear gauge condition given by equation (\ref{46}), the result follows from the following arguments.  Starting from a field configuration $A^{a}_{\mu}$ that does not satisfy equation (\ref{46}), we can always gauge transform to one that satisfies the non-linear gauge.  This field configuration is given by
\begin{equation}\label{54}
A^{a}_{\mu}(x)\vert_{\mathcal{F}_{_{o\cdots o}}}=A^{a}_{\mu}(x)-D^{ab}_{\mu}(x)\int d^{4}x'H^{bc}(x,x';A)[(\partial\cdot D)^{cd}(\partial\cdot A^{d})]_{x'},
\end{equation}
where $H^{ab}(x;x'; A)$ is the Green function of the non-singular operator
\begin{equation}\label{55}
\mathbf{\Theta}^{ab}=(D\cdot\partial)^{ac}(\partial\cdot D)^{cb}-\epsilon^{acd}[\partial(\partial\cdot A^{c})]\cdot D^{db}.
\end{equation}
The non-singularness of $\mathbf{\Theta}^{ab}$, even if $(\partial\cdot D)$ has a zero mode $(\partial\cdot A^{a}=\mathit{f}^{a}\neq 0)$ is verified in first-order perturbation theory.  Since this is crucial to what follows, we will outline the proof of this claim.

First, $\mathbf{\Theta}^{ab}$ is hermitian on the submanifold defined by equation (\ref{46}).  Since the first terms of $\mathbf{\Theta}$ is a fourth-order operator (dominant term), with zero mode $\partial\cdot A$, the zero mode of $\mathbf{\Theta}$, if it exists, must be of the form
\begin{equation}\label{56}
{z}^{a}=\partial\cdot A^{a}+\lambda^{a},
\end{equation}
with $\lambda^{a}\ll\partial\cdot A^{a}$.  The correction $\lambda^{a}$ must be solved from
\begin{equation}\label{57}
(D\cdot\partial)(\partial\cdot D)\lambda = [\partial(\partial\cdot A)\cdot D](\partial\cdot A).
\end{equation}
The solution to equation (\ref{57}) only exists if the zero mode $\partial\cdot A$ is orthogonal to the source in the above equation, i.e.
\begin{equation}\label{58}
\int d^{4}x(\partial\cdot A^{a})\epsilon^{abc}[\partial(\partial\cdot A^{b})\cdot D^{cd}](\partial\cdot A^{d})= 0.
\end{equation}
But by integration by parts, it is easy to show that the above integral is
\begin{equation}\label{59}
\geq c^{2}\parallel\partial\cdot A^{a}\parallel^{2},
\end{equation}
where $c^{a}$ is the minimum value of $\partial\cdot A^{a}$.  Since equation (\ref{58}) can never be satisfied, $\lambda$ does not exist and $\mathbf{\Theta}$ is non-singular.

Going back to equation (\ref{54}), we find that
\begin{eqnarray}
dA^{a}_{\mu}\vert_{\mathcal{F}_{o\cdots o}}&=& dA^{a}_{\mu}(x)-\epsilon^{abe} dA^{e}_{\mu}(x)\int d^{4}y H^{bc}(x;y;A)[(D\cdot\partial)^{cd}(\partial\cdot A^{d})]_{y}\nonumber\\
&-& D^{ab}_{\mu}(x)\int d^{4} y H^{bc}(x,y;A)[(D\cdot\partial)^{cd}\partial_{\alpha}d A^{d}_{\alpha}]_{y}\nonumber\\
&-&D^{ab}_{\mu}(x)\int d^{4}y H^{bc}(x,y;A)[(-)\epsilon^{cde} dA^{e}_{\alpha}\partial_{\alpha}(\partial\cdot A^{d})]_{y}\nonumber\\
&-& D^{ab}_{\mu}(x)\int d^{4} y d^{4} z\frac{\delta H^{bc}}{\delta A^{f}_{\alpha}(z)}(x,y;A)dA^{f}_{\alpha}(z)[(D\cdot\partial)^{cd}(\partial\cdot A^{d})]_{y}\label{60}
\end{eqnarray}
where $H^{ab}$ is the Green function of $\mathbf{\Theta}$.  The last term is evaluated by using
\begin{equation}\label{61}
\delta H = -\int H\delta\mathbf{\Theta} H.
\end{equation}

Substituting equation (\ref{60}), in equations (\ref{52}), we verify equation (\ref{53}) after doing integration by parts.\\

At this point we ask, do the submanifolds $\mathcal{F}_{_{c_{I}}\cdots c_{\mathcal{H}}}$ given by equations (\ref{42}) to (\ref{48}) foliate the entire configuration space of Yang-Mills theory in the same way that the Coulomb-like surfaces discussed in Section 3 foliate the Abelian configuration space?  The answer is no as we will argue below.\\

In Section 3, the foliation by Coulomb-like surfaces of the Abelian configuration space is not subject to any restrictions.  Thus, the entire configuration space can be foliated by the Coulomb-like surfaces, which are derived from the condition of orthogonality of the orbit to the submanifold.\\

On the other hand, equations (\ref{41}) to (\ref{61}) have built-in conditions.\\

First, the non-linear gauge given by equation (\ref{46}) has two
regimes:
\begin{equation}\label{62}
\partial\cdot A^{a} = 0,
\end{equation}
\begin{equation}\label{63}
\partial\cdot A^{a} = f^{a}\neq 0;\, (\partial\cdot D)^{ab}f^{b}=0.
\end{equation}

The first is the Coulomb gauge, which describes the physical degrees of freedom in the short-distance regime as argued in reference \cite{Essay}.  The second is the quadratic regime, which was shown to yield non-perturbative physics in the $f^{a}$ dynamics (see references \cite{linear},\cite{chaos}).\\
Second, field configurations satisfying $\partial\cdot A^{a}\neq 0$, with $det(\partial\cdot D)\neq 0$, can be gauge transformed to the Coulomb surface.  For these field configurations, the orbit, although not necessarily orthogonal to Coulomb-like surfaces, will never be tangential to these surfaces.  This is seen by computing the angle between the tangent to the orbit and normal to the surface given by
\begin{eqnarray}\label{64}
cos\theta_{c}\sim\int d^{4}x(D^{ab}_{\mu}\Lambda^{b})\partial_{\mu}\Lambda^{a}\nonumber\\
\sim\int d^{4}x\Lambda^{a}(\partial\cdot D)^{ab}\Lambda^{b}.
\end{eqnarray}
Since $(\partial\cdot D)$ is non-singular, it has no zero modes and $\theta_{c}$ is never equal to $\frac{\pi}{2}$.  This means the orbit is never tangential to the Coulomb-like surface.\\

Third, if $det(\partial\cdot D)=0$, then $(\partial\cdot D)$ has zero mode $z^{a}(x)$ and $cos\theta_{c}$ may be zero.  In this case, the orbit is tangential to the Coulomb-like surface.\\

Fourth, if the zero mode of $(\partial\cdot D)$ is $\partial\cdot
A^{a}$, then the appropriate submanifolds are defined by equations
(\ref{42}) to (\ref{61}).  However, it must be noted that the
submanifolds $\mathcal{F}_{_{c_{I}}\cdots c_{\mathcal{H}}}$ are
only valid in the regimes defined by the restricting conditions.
And these are, for $\mathcal{F}_{_{c_{I}}\cdots c_{\mathcal{H}}}$,
(a)\,$\partial\cdot A^{a}\neq 0$, (b) the only zero mode of
$(\partial\cdot D)$ is $\partial\cdot A^{a}$, which leads to the
non-singular character of the fourth-order operator $\Theta^{ab}$.
The important thing though is that the orbit, although not
orthogonal to the submanifold, is also never tangential to it (see
figure 3).  This follows from
\begin{eqnarray}\label{45}
cos\theta\sim\int d^{4}x(D^{ab}_{\mu}\Lambda^{b})\frac{\delta F^{\Lambda}}{\delta A^{a}_{\mu}(x)},\nonumber\\
\sim\int d^{4}x\Lambda^{a}\Theta^{ab}\Lambda^{b},
\end{eqnarray}
which is never zero because $\Theta^{ab}$ has no zero modes.\\

Finally, we conjecture that maybe the non-existence of submanifolds which can foliate the entire configuration space is a reflection of the fact that the physical degrees of freedom of Yang-Mills theory change with the distance scale.  In short-distance regime, massless, transverse gluons are valid degrees of freedom.  In the long-distance regime, transverse gluons are not valid degrees of freedom.  This paper and references \cite{linear} and\cite{chaos} argue that the scalar $f^{a}$ and vector fields $t^{a}_{\mu}$, which arise from the non-linear gauge should be used instead.  For comparison purposes, note that the Coulomb-like submanifolds, which comes from orthogonality condition, foliate the entire Abelian configuration space and this may be related to the fact that the transverse photon is a physical degree of freedom in all distance scales.

\begin{figure}
\centering
\includegraphics[totalheight=2.5in]{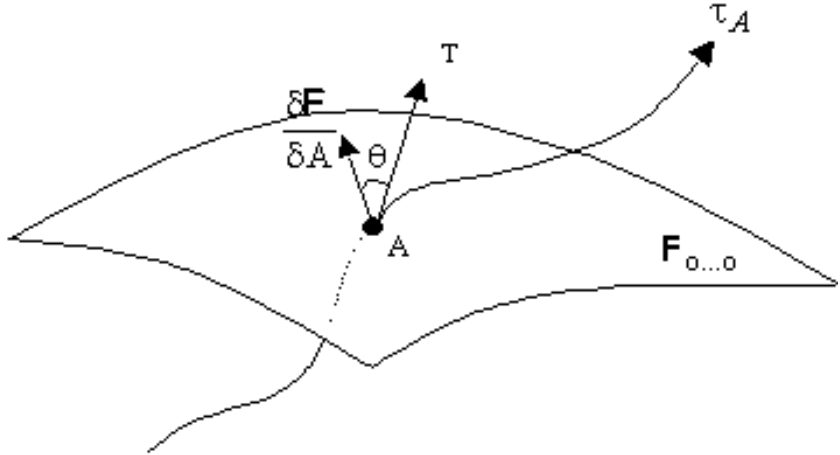}
\caption{The gauge-fixing surface $\mathcal{F}_{o\cdots o}$, defined by the non-linear gauge and the orbit $\tau_{A}$ make an angle $\theta$, which is never equal to $\frac{\pi}{2}$. $\theta$ is the angle between the normal to $\mathcal{F}_{o\cdots o}$ (given by $\frac{\delta\mathcal{F}_{o\cdots o}}{\delta A}$) and the tangent to the orbit (given by $\mathcal{T}$).}
\label{fig3}
\end{figure}

\section{Conclusion}

In this paper, we have established the geometrical basis of the non-linear gauge condition.  We have shown that although the orbit is never orthogonal to $\mathcal{F}_{o\cdots o}$, it is also never tangential to the surface $(\theta_{c}\neq\frac{\pi}{2})$.  Unfortunately, the submanifolds $\mathcal{F}_{_{c_{I}}\cdots c_{\mathcal{H}}}$ is shown to be not capable of foliating the entire configuration space but only the non-perturbative regime of Yang-Mills theory.

\section{Acknowledgement}

In the early stages, this research was supported in part by the National Research Council of the Philippines.  The later stage of this research was supported by the Natural Sciences Research Institute and the University of the Philippines System.  The research was started when the author visited the University of Mainz through the financial support of the Alexander von Humboldt Stiftung.  Discussions with Martin Reuter are gratefully acknowledged.

\end{document}